\documentclass[12pt]{iopart}

\usepackage{bm}
\usepackage{graphicx}

\begin{document}

\title{Analytical approximations to the spectra of quark-antiquark potentials}

\author{Paolo Amore}
\ead{paolo@ucol.mx}
\address{Facultad de Ciencias, Universidad de Colima,
  Bernal D\'{i}az del Castillo 340, Colima, Colima, Mexico}

\author{Arturo De Pace}
\address{Istituto Nazionale di Fisica Nucleare, Sezione di Torino, via
  Giuria 1, I-10125 Torino, Italy}

\author{Jorge Lopez}
\address{Physics Department, University of Texas at El Paso, El Paso,
  Texas, USA} 

\begin{abstract}
    
A method, recently devised to obtain analytical approximations to certain
classes of integrals, is used in combination with the WKB expansion to derive 
accurate analytical expressions for the spectrum of quantum potentials.
The accuracy of our results is verified by comparing them both with the
literature on the subject and with the numerical results obtained with a
Fortran code. 
As an application of the method that we propose, we consider the meson
spectroscopy with various phenomenological potentials.

\end{abstract}

\pacs{02.30.Mv,03.65.Ge,45.10.Db}

\date{\today}

\maketitle

\section{Introduction}

Recently, in Refs.~\cite{AS:04,AAFS:04} a method has been developed that 
allows one to obtain analytical approximations with arbitrary precision to a large class 
of integrals appearing in many diverse physical problems. 
The method is based on a carefully chosen expansion, which has been shown to
converge uniformly and gives excellent results already at first order. 
In Refs.~\cite{AS:04,AAFS:04} such method has been applied to a few problems in
Classical Mechanics and General Relativity, allowing one to obtain simple
analytical formulas that work even in the non-perturbative regime. 
In this paper we consider a further application of the method, showing that
it is possible to obtain accurate {\em analytical} expressions for the spectrum
of a quantum potential by using it in conjunction with the WKB expansion. 

WKB is a semi-classical approximation, whose accuracy improves for large
quantum numbers, based on an expansion in $\hbar$ of the Schr\"odinger
equation: it has been widely used in the past to get approximate evaluations of
the spectra of quantum potentials and it is generally employed at the leading
order in $\hbar$, although its extension at higher orders has allowed one 
to recover, in special cases,  the exact energy eigenvalues \cite{BOW:77}. 
On the other hand, for a generic potential function $V(x)$ the calculation of
the WKB correction to a given order can be done only numerically and, in
general, no analytical expression can be found for the WKB integrals and, as a
consequence, for the spectrum.
 
The WKB method has already been employed also in the specific case of the
phenomenological quark-antiquark potentials, using both non-relativistic 
\cite{Brau00} and relativistic \cite{Cea82} kinematics, showing a remarkably 
good accuracy even for the lowest quantum numbers. However, the solution of the
WKB equation, although simpler than the original Schr\"odinger equation, has
been again done numerically and only in the limit of large quantum numbers it
has been possible to obtain the spectrum in a closed form.

In this paper, by applying the method of Ref.~\cite{AS:04} to the calculation
of the WKB integrals we show that one can obtain an accurate {\em analytical}
approximation to the {\em full spectrum} of the potential (see also
Ref.~\cite{See83}).  
The accuracy of the approximation depends both on the order to which the WKB
expansion is considered and on the order to which our method is applied.
We apply the method to a few phenomenological quark-antiquark potentials that
have been used in the literature to calculate the spectra of heavy mesons.

In Sect.~\ref{sec:formalism} we discuss our method and in
Sect.~\ref{sec:application} we first apply it to the case of the 
non-relativistic Cornell potential \cite{Eic75} and then to the case of a
linear potential with relativistic kinematics.
Finally, in Sect.~\ref{sec:concl} we briefly summarize our results.

\section{Formalism}
\label{sec:formalism}

The standard WKB approximation is obtained by solving for the energy the
following equation:
\begin{equation}
  \int_{x_-}^{x_+}dx\, \sqrt{E-V(x)} = \pi(n+\frac{1}{2}),
\end{equation}
where $x_\pm$ are the classical turning points of the potential $V(x)$ and $E =
V(x_\pm)$ is the energy. 

Let us concentrate on the integral 
\begin{equation}
  {\mathcal J} = \int_{x_-}^{x_+}dx \, \sqrt{E-V(x)}.
\end{equation}
In the spirit of the Linear Delta Expansion \cite{lde} we interpolate the
potential $V(x)$ as 
\begin{eqnarray}
  V_\delta(x) = V_0(x) + \delta (V(x)-V_0(x)),
\label{EQN1}
\end{eqnarray}
where $V_0(x)$ is a suitably chosen potential that depends on one or more
arbitrary parameters $\lambda_1, \lambda_2, ...$ that we collectively call
$\lambda$ in the following.
$V_\delta(x)$ reduces to the full potential for $\delta=1$.
We want to perform an expansion in $\delta$ without moving the inversion
points: for this reason we impose that $x_\pm$ be the inversion points also of
the potential $V_0(x)$.
As a result, the energy $E_0$ that the particle would possess if it were moving
only in the potential $V_0(x)$ is given by $E_0 = V_0(x_\pm)$.

We therefore introduce a new integral
\begin{equation}
  {\mathcal J}^{(\delta)} = \int_{x_-}^{x_+}dx \, \sqrt{E_0-V_0(x)} 
    \left[ 1 + \delta \Delta(x) \right]^{1/2},
\end{equation}
where
\begin{equation}
  \Delta(x) \equiv \frac{E -E_0 - V(x) + V_0(x)}{E_0-V_0(x)}.
\end{equation}
Of course, one has ${\mathcal J}^{(\delta=1)}\equiv{\mathcal J}$.

We then treat the term proportional to $\delta$ as a perturbation and perform
an expansion in $\delta$,
\begin{equation}
  {\mathcal J}^{(\delta)} = \sum_{n=0}^\infty \delta^n {\mathcal J}_n.
\end{equation}
Such expansion will converge uniformly when $|\Delta(x)| < 1$ in the 
region $x_- \leq x \leq x_+$ \cite{AS:04}. This condition selects a particular
region in the parameter space $\lambda$; however, maximal convergence is
achieved when the Principle of Minimal Sensitivity (PMS) \cite{Ste81} is used, 
i.~e. when, given the $N$-th order approximation to ${\mathcal J}$,
\begin{equation}
\label{eq:approximant}
  {\mathcal J}_{(N)} = \sum_{n=0}^N {\mathcal J}_n,
\end{equation}
the condition 
\begin{equation}
\label{eq:PMS}
  \partial{\mathcal J}_{(N)}/\partial\lambda = 0
\end{equation}
is enforced. Eq.~(\ref{eq:PMS}) fixes the parameters $\lambda$, which are then
inserted back in Eq.~(\ref{eq:approximant}) yielding the approximant of 
${\mathcal J}$. 

Notice that if the potential $V_0(x)$ is chosen appropriately it will be
possible to calculate analytically each term in the expansion.  

\section{Applications: meson spectroscopy}
\label{sec:application}

\subsection{Non-relativistic kinematics: the Cornell potential}
\label{subsec:Cornell}

Following \cite{Brau00} we write
\begin{eqnarray}
  \int_{r_-}^{r_+} \sqrt{2 \mu \left[ E + \frac{\kappa}{r} - a r - \frac{L^2}{2
    \mu r^2}\right]} \ dr =  \pi \left( n + \frac{1}{2} \right)
\end{eqnarray}

The effective potential in this case is given by
\begin{equation}
\label{eq:Corneff}
  V_{{\rm eff}}(r) = - \frac{\kappa}{r} + a r + \frac{L^2}{2 \mu r^2}
\end{equation}
where the last term is the centrifugal barrier and $r_{\pm}$ are the classical
inversion points. 

\begin{figure}
\begin{center}
\includegraphics[width=9cm]{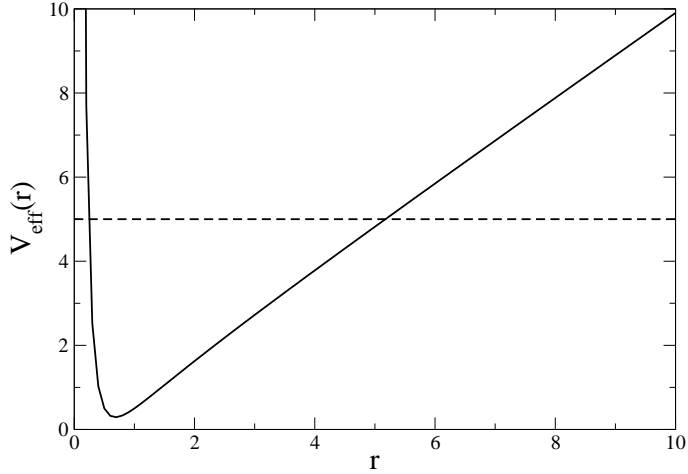}
\caption{Effective potential for $a=\kappa=L=\mu=1$.
\label{fig:figpot}} 
\end{center}
\end{figure}

As one can see from Fig.~\ref{fig:figpot}, this potential is not symmetrical
with respect to the minimum, which we call $r_0$. We therefore write
\begin{eqnarray}
\label{eq:J}
\fl
  {\mathcal J} &\equiv& \int_{r_-}^{r_+} dr \sqrt{2 \mu \left[ E +
    \frac{\kappa}{r} - a r - \frac{L^2}{2 \mu r^2}\right]} \nonumber \\
\fl
  &=&  \int_{r_-}^{r_0} dr \sqrt{2 \mu \left[ E + \frac{\kappa}{r} - a r -
    \frac{L^2}{2 \mu r^2}\right]}   
    + \int_{r_0}^{r_+} dr \sqrt{2 \mu \left[ E + \frac{\kappa}{r} - a r -
    \frac{L^2}{2 \mu r^2}\right]} \nonumber \\ 
\fl
  &\equiv& {\mathcal J}_+ + {\mathcal J}_- 
\end{eqnarray}
and evaluate the two integrals separately. 

Following the procedure outlined in the previous section, we can write
\begin{eqnarray}
  {\mathcal J}_{\pm} &=&  \mp \int_{r_{\pm}}^{r_0} dr \sqrt{2 \mu \left[ E +
      \frac{\kappa}{r} - a r - \frac{L^2}{2 \mu r^2}\right]} \nonumber \\ 
  &=& \mp \int_{r_{\pm}}^{r_0} dr \sqrt{2 \mu \left[ E_{0\pm} -
     V_{0\pm}(r)\right]} \sqrt{1+ \delta \Delta_{\pm}(r)} , 
\end{eqnarray}
where
\begin{equation}
  \Delta_{\pm}(r) \equiv \frac{E -E_{0\pm} + 
    \frac{\displaystyle \kappa}{\displaystyle r} - a r - 
    \frac{\displaystyle L^2}{\displaystyle 2 \mu r^2}+ V_{0\pm}(r)}
    { E_{0\pm} - V_{0\pm}(r)} 
\end{equation}
and
\begin{equation}
  V_{0\pm}(r) = \frac{\lambda_{\pm}^2}{2} (r-r_0)^2, 
    \quad E_{0\pm}=V_{0\pm}(r_{\pm}), 
\end{equation}
$\lambda_{\pm}$ being arbitrary parameters.

Provided that $\left|\Delta_{\pm}(r)\right|< 1$ for all $r \in (r_{\pm},r_0)$, 
one can expand the expression for ${\mathcal J}_{\pm}$ in powers of
$\Delta_{\pm}(r)$.
At first order one obtains
\begin{equation}
\label{eq:J1pm}
  {\mathcal J}_{(1)\pm} = \sqrt{\mu}\lambda_{\pm} \left(
    \frac{1}{2}{\mathcal H}_{0\pm} + 
    \frac{{\mathcal H}_{1\pm}}{\lambda_{\pm}^2} \right),
\end{equation}
where 
\begin{equation}
  {\mathcal H}_{1\pm} = \pm \left[ a {\mathcal H}_{1\pm}^a + 
    \frac{1}{r_{\pm}}\left(\kappa-\frac{L^2}{2\mu r_{\pm}}\right) 
    {\mathcal H}_{1\pm}^b - \frac{L^2}{2\mu r_{\pm}} {\mathcal H}_{1\pm}^c 
    \right],
\end{equation}
${\mathcal H}_{0\pm}$, ${\mathcal H}_{1\pm}^a$ ,${\mathcal H}_{1\pm}^b$ and
${\mathcal H}_{1\pm}^c$ being analytic functions of $r_{\pm}$ and $r_0$ given
in the Appendix. 

\begin{figure}
\begin{center}
\includegraphics[width=9cm]{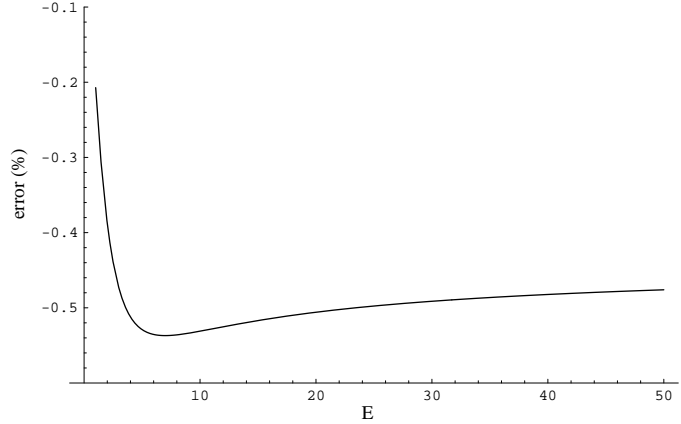}
\caption{\% error of ${\mathcal J}_{(1)+}+{\mathcal J}_{(1)-}$ for 
  $a=\kappa=L=\mu=1$ as a function of $E$}
\label{fig:Icornell}
\end{center}
\end{figure}

By minimizing ${\mathcal J}_{(1)\pm}$ with respect to the corresponding
parameter $\lambda_{\pm}$ one finds the optimal values,
\begin{equation}
  \lambda_{\pm} = \sqrt{\frac{2 {\mathcal H}_{1\pm}}{{\mathcal H}_{0\pm}}}, 
\end{equation}
and, as a consequence, the optimal first order approximation to the integrals
in Eq.~(\ref{eq:J}) turns out to be
\begin{equation}
  {\mathcal J}_{(1)\pm} = \sqrt{2\mu} \sqrt{{\mathcal H}_{0\pm}
    {\mathcal H}_{1\pm}}.
\end{equation}
In Fig.~\ref{fig:Icornell} one can see the typical accuracy one gets using this
approximation.

One has now to impose the WKB condition
\begin{equation}
  {\mathcal J}_{(1)+} + {\mathcal J}_{(1)-} = \pi\left(n+\frac{1}{2}\right)
\end{equation}
and solve for $E\equiv E_n$. 
In order to simplify this equation we expand the left hand side in $1/E_n$ to
leading order: by inspecting the resulting expression one can easily check that
the WKB condition can be satisfied by expanding the energy as a series in 
$1/n$, i.~e. assuming the following ansatz
\begin{equation}
\label{eq:EnCorn}
  E_n = \sum_{n=0}^\infty c_k \left(n+1/2+\frac{L (8+17 \pi )}{32 \sqrt{2} \pi}
    \right)^{2/3-k/6}
\end{equation}
and determining the coefficients $c_k$ so that the quantization condition is 
obeyed. These coefficients can be recursively calculated using any computer  
albegra program: stopping at order 10 in the $1/n$ expansion one finds
\numparts
\label{eq:cn}
\begin{eqnarray}
\fl
  c_0 = \frac{2^{2/3} a^{2/3} \sqrt[3]{\pi }}{\sqrt[3]{\mu (\pi -2) }} \\
\fl
  c_1 = c_2 = c_3 = 0 \\
\fl
  c_4 = a r_0\left(1-\frac{2}{3} \sqrt{\frac{\pi }{\pi-2}}\right) \\
\fl
  c_5 = \frac{a \left(\frac{\displaystyle \pi }{\displaystyle \pi-2}
    \right)^{5/12} \sqrt{L r_0}}{3 2^{11/12} \sqrt[6]{a \mu }} \\ 
\fl
  c_6 = 0 \\
\fl
  c_7 = -\frac{\sqrt{a} \left[(3\pi-8) L^2+40 \kappa \mu \pi r_0\right]}{96 
    \sqrt[4]{\pi-2} (2 \pi )^{3/4} \sqrt{L \mu r_0}} \\ 
\fl
  c_8 = \frac{\sqrt[3]{a}}{4608 \mu ^{2/3} (2 (\pi-2))^{2/3} \pi^{11/6} r_0}
    \left\{3 \sqrt{\pi-2} (64+21 (\pi-16 ) \pi ) L^2 \right. \nonumber \\
   \left. +32 \mu \pi r_0 \left[3 \kappa \left[(8-21 \pi )
    \sqrt{\pi-2}-48 (\pi-2) \sqrt{\pi }\right] \right. \right. \nonumber \\
   \left. \left.  -4 a \left[24 
    \sqrt{\pi-2}+(18-13 \pi) \sqrt{\pi}\right] r_0^2\right]\right\} \\ 
\fl
  c_9 = \frac{\sqrt[6]{a}}{18432 L^{3/2}\mu^{5/6}(2(\pi-2))^{7/12}\pi^{23/12}
    r_0^{3/2}} \left[-256 a L^2 \mu \pi \left(3 \sqrt{\pi-2}(4-3\pi) 
    +2 \pi ^{3/2}\right) r_0^3 \right. \nonumber \\
   \left. -3 \sqrt{\pi-2 } 
    \left((-192+\pi (112+97 \pi )) L^4-16 \kappa \mu \pi (77 \pi-24 ) r_0 
    L^2 \right. \right. \nonumber \\
    \left. \left. +448 \kappa ^2 \mu ^2 \pi ^2 r_0^2\right)\right]  \\ 
\fl
  c_{10} = \frac{1}{49152 \sqrt{2} L \mu  \pi ^3 r_0^2}\left\{
    (8+\pi)[64+\pi  (181\pi-144)] L^4+32 \mu \pi r_0 
    \left[\kappa  [64-3 \pi  (32+3 \pi )] \right. \right. \nonumber \\
   \left. \left. \quad -16 a [16+\pi  (13\pi-66)] 
    r_0^2\right] L^2+256 \kappa ^2 \mu ^2 \pi ^2 (8+25 \pi )
    r_0^2 \right\}.
\end{eqnarray}
\endnumparts
In Table~\ref{tab:Corn} we compare the analytic spectrum of the $c\bar{c}$
and $b\bar{b}$ systems --- obtained from Eqs.~(\ref{eq:EnCorn}) and
(\ref{eq:cn}) --- to the spectra obtained from the exact solutions of the
Schroedinger and WKB equations, respectively \cite{Brau00}.

\begin{table}[!htb]
\caption{\label{tab:Corn} Masses (in MeV) of the lowest $c\bar{c}$ and
  $b\bar{b}$ states from the exact solutions of the Schroedinger and WKB
  equations \protect\cite{Brau00} and from the analytical approximation of
  Eqs.~(\ref{eq:EnCorn}) and (\ref{eq:cn}). Quark masses and potential
  parameters are taken from Ref.~\cite{Ful94}. }
\begin{indented}
\item[]\begin{tabular}{llll}
 \br
 States & Exact & WKB$_{{\rm exact}}$ & WKB$_{{\rm analytic}}$ \\
 \br
 $c\bar{c}$ \\
 1S & 3067 & 3062 & 3097 \\
 2S & 3693 & 3691 & 3722 \\
 1P & 3497 & 3497 & 3463 \\
 2P & 3991 & 3991 & 3980 \\
 1D & 3806 & 3806 & 3791 \\
 2D & 4242 & 4242 & 4235 \\
 \br
 $b\bar{b}$ \\
 1S &  9448 &  9439 &  9705 \\
 2S & 10007 & 10003 & 10131 \\
 1P &  9901 &  9900 &  9865 \\
 2P & 10261 & 10261 & 10255 \\
 1D & 10148 & 10148 & 10098 \\
 \br
\end{tabular}
\end{indented}
\end{table}

As one can see from the table, in the case of the $c\bar{c}$ mesons the error
in the analytic expressions is at most 1\% for the lowest state and it is
rapidly decreasing with the excitation energy; in the case of the $b\bar{b}$
mesons one goes from a 2.8\% error in the ground state up to a 0.06\% error for
the 2P state. Note that the accuracy of the analytic approximation can be
arbitraly improved simply by extending the order in the $1/E_n$ and $1/n$
expansions and that similar analytic formulae can also be obtained for any
observable by expanding its WKB expression.

It is also quite straightforward to get simple formulae in the asymptotic
regimes. For instance, if we assume $n \gg L \gg 1$, the energy is approximated
by our formula as 
\begin{equation}
  E_n \approx c_0 n^{2/3} \approx 2.2245 \frac{a^{2/3}}{\mu^{1/3}} n^{2/3}, 
\end{equation}
which is extremely close to the result of Eq.~(10) of Ref.~\cite{Brau00}:
\begin{equation}
  E_n \approx 2.2309 \frac{a^{2/3}}{\mu^{1/3}} n^{2/3}.
\end{equation}
On the other hand, in the regime $L\gg n\gg 1$  we have to make explicit the
dependence of $r_0$ upon $L$ and then expand around $L=\infty$; 
we obtain:
\begin{equation}
  E_n \approx 1.62176 \frac{a^{2/3}}{\mu^{1/3}} L^{2/3},
\end{equation}
which again can be compared with the formula (9) of Ref.~\cite{Brau00}:
\begin{equation}
  E_n \approx 1.5 \frac{a^{2/3}}{\mu^{1/3}} L^{2/3}.
\end{equation}

\subsection{Relativistic kinematics: linear potential}

A generalization of the Schr\"odinger equation to include relativistic effects
can be done by starting with the Bethe-Salpeter equation and making a few
approximations, namely assuming an instantaneous local potential and neglecting
spin and the coupling of ``large'' to ``small'' components in the relativistic
wave function. 
The resulting equation --- also known as spinless Salpeter equation --- has
exactly the form of a Schr\"odinger equation with the kinetic term replaced
with $\sqrt{-\bm{\nabla}^2+m_1^2} + \sqrt{-\bm{\nabla}^2+m_2^2}$.

\begin{figure}
\begin{center}
\includegraphics[width=9cm]{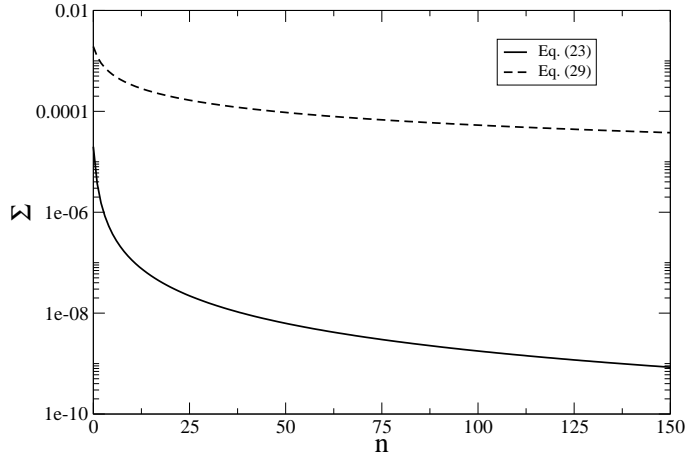}
\caption{Error over the spectrum defined as $\Sigma \equiv
  (E_n^{\mathrm exact}-E_n^{\mathrm approx})/E_n^{\mathrm exact}$ 
  using the analytical formulas obtained and taking $m=\mu=1$ and $l=0$.
\label{fig:error_Em1eqm2}} 
\end{center}
\end{figure}

Although the relativistic Schr\"odinger equation has been used in the
literature with many realistic phenomenological potentials, in the following we
shall consider for simplicity the linear potential $V(r)=\mu^2 r$.
In the relativistic case the WKB condition turns out to be slightly modified
and it can be written as \cite{Cea82}
\begin{equation}
\label{eq:WKBl}
  {\mathcal J} = \pi\left( n + \frac{l}{2} + \frac{3}{4} \right),
\end{equation}
where
\begin{equation}
\label{eq:JWKBl}
  {\mathcal J} = \int_0^{r_+}dr \, \sqrt{\frac{(E-\mu^2 r)^2}{4} - 
    \frac{m_1^2+m_2^2}{2} + \frac{(m_1^2-m_2^2)^2}{4(E-\mu^2 r)^2} },
\end{equation}
with 
\begin{equation}
  r_+ \equiv \frac{E-m_1-m_2}{\mu^2} \ .
\end{equation}
In the limit $m_1 = m_2 = m$, Eq.~(\ref{eq:WKBl}) can be integrated exactly to
give \cite{Cea82} 
\begin{equation}
  \frac{E}{2} \sqrt{\frac{E^2}{4} - m^2} - m^2 \log \left[\frac{E}{2 m} +
  \sqrt{\frac{E^2}{4 m^2} - 1}  
  \right]= \mu^2 \pi \left( n + \frac{l}{2} + \frac{3}{4}\right) .
\label{eq_1a}
\end{equation}
Before dealing with the general case $m_1\neq m_2$, we first want to show that
Eq.~(\ref{eq_1a}) can be used to obtain an analytical approximation to the
spectrum.
We make the ansatz:
\begin{equation}
\fl
  E = \sum_{k=0}^\infty \left\{ c_k \left(n +\frac{l}{2} +
    \frac{3}{4}\right)^{1/2-k} + 
    d_k \left(n +\frac{l}{2} + \frac{3}{4}\right)^{-1/2-k} 
    \log \left(n +\frac{l}{2} + \frac{3}{4}\right) \right\}.
\label{eq_1b}
\end{equation}
The unknown coefficients $c_k$ and $d_k$ in this expansion are determined by 
inserting Eq.~(\ref{eq_1b}) into Eq.~(\ref{eq_1a}) and by Taylor expanding in
$1/(n+l/2+3/4) \ll 1$. 
\begin{eqnarray}
\fl
  E_{n,l} &\approx& \frac{m^2 + {\mu}^2 \left( 3 + 2l + 4n \right) \pi +
    m^2 \log \left(\frac{\displaystyle 4m\pi}{\displaystyle \mu}\right)} 
    {\mu {\sqrt{3 + 2l + 4n}}{\sqrt{\pi}}} \nonumber \\ 
\fl
  && + \frac{m^2 \log \left(\frac{\displaystyle 3}{\displaystyle 4} + 
    \frac{\displaystyle l}{\displaystyle 2} + n \right) \left[ {{m}}^2 +
    {\mu}^2 \left( 3 + 2l + 4n \right) \pi - m^2 \log \left(
    \frac{\displaystyle 4m\pi }{\displaystyle \mu}\right) \right]}
    {{\mu}^3 {\left( 3 + 2l + 4n \right) }^{3/2} {\pi }^{3/2}} + \dots . 
\label{eq_1c}
\end{eqnarray}
This formula has been obtained truncating the sum into Eq.~(\ref{eq_1b}) to 
$k=2$. In Fig.~\ref{fig:error_Em1eqm2} we compare the numerical solutions
of Eq.~(\ref{eq:WKBl}) with the analytical approximation of Eq.~(\ref{eq_1c}), 
using $m=\mu=1$ and $l=0$ (solid curve). As one can see the accuracy is very
good. 

We now want to extend our previous results to cases where the quantization
integral cannot be performed analytically, such as for $m_1 \neq m_2$. 
Following the procedure outlined in Sect.~\ref{sec:formalism} we can rewrite 
Eq.~(\ref{eq:JWKBl}) as
\begin{eqnarray}
\fl
  {\mathcal J}^{(\delta)} = \int_0^{r_+}dr \, 
    \sqrt{\frac{(E_0-\lambda^2 r)^2}{4} + \delta \left[ 
    \frac{(E-\mu^2 r)^2}{4}- \frac{m_1^2+m_2^2}{2} + 
    \frac{(m_1^2-m_2^2)^2}{4(E-\mu^2 r)^2} - \frac{(E_0-\lambda^2 r)^2}{4} 
    \right]}, \nonumber \\ 
\label{eq_1d}
\end{eqnarray}
where $\lambda$ is an arbitrary parameter and $\delta$ is used for power
counting. For $\delta = 1$ the expression above reduces to the one in 
Eq.~(\ref{eq:WKBl}). $E_0$ is fixed by asking that the zero-th order term in 
$\delta$ vanishes at the inversion point and its value is
\begin{equation} 
  E_0 = \lambda^2 r_+ .
\end{equation}
If we expand Eq.~(\ref{eq_1d}) in powers of $\delta$ at order $N$ and then set 
$\delta=1$, we obtain an expression that depends upon $\lambda$. 
We again take the approximant ${\mathcal J}_{(N)}$ and impose 
$d{\mathcal J}_{(N)}/d\lambda=0$ to minimize such dependence.
To leading order in $\delta$ one obtains
\begin{equation}
\fl
  \lambda^4 = \mu^4 \frac{ -8\,{(\Delta m)}^2 \left( E + 2\overline{m} \right)
    + E \left( E - 2\overline{m} \right) \left( E + 6\overline{m} \right) + 8 
    {(\Delta m)}^2 E \log(2E\overline{m})}{E{\left(E-2\overline{m}\right)}^2}, 
\label{eq_1e}
\end{equation}
where $\Delta m = (m_1-m_2)/2$ and $\overline{m} = (m_1+m_2)/2$.

\begin{figure}
\begin{center}
\includegraphics[width=9cm]{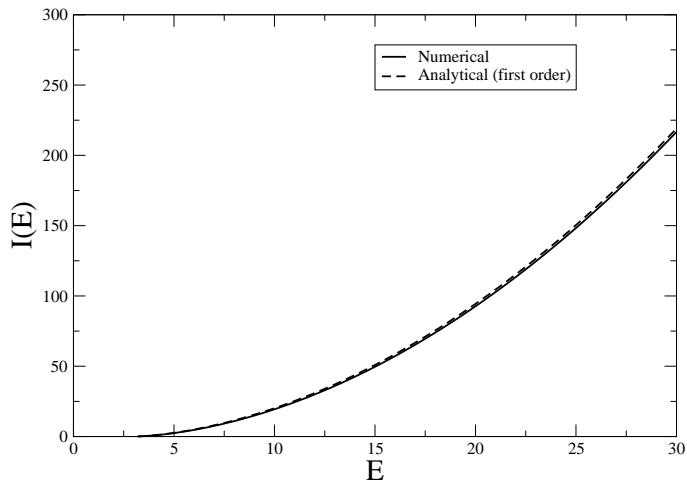}
\caption{Comparison between the exact numerical value and the analytical 
  approximation at first order (Eq.~(\protect\ref{eq_1f})) of the quantization
  integral ${\mathcal J}$, taking $m_1 = 1$, $m_2=2$ and $\mu=1$. 
\label{fig:Jm1nem2}}
\end{center}
\end{figure}

Thus, to order $\delta$ we find 
\begin{eqnarray}
  {\mathcal J}_{(1)} &=& \frac{ E - 2\overline{m}}{4 {\sqrt{E}} {\mu}^2}
    \left\{ E \left( E - 2\overline{m} \right) \left( E + 6\overline{m} 
    \right) - 8 {(\Delta m)}^2 \left[ E + 2\overline{m} - E\log(2) \right] 
    \right. \nonumber \\
  && \left. + 8 {(\Delta m)}^2 E \left[ \log (E) + \log (\overline{m}) \right]
    \right\}^{1/2}.
\label{eq_1f}
\end{eqnarray}
In Fig.~\ref{fig:Jm1nem2} we compare the numerical result for ${\mathcal J}$
obtained using $m_1 = 1$, $m_2=2$ and $\mu=1$, with the one 
of Eq.~(\ref{eq_1f}), as a function of the energy $E$. Eq.~(\ref{eq_1f})
provides a quite simple and precise approximation to the exact integral.

By introducing the ansatz 
\begin{equation}
\fl
  E = \sum_{k=0}^\infty \left\{ c_k \left(n +\frac{l}{2} +
    \frac{3}{4}\right)^{1/2-k/2} +  d_k \left(n +\frac{l}{2} + 
    \frac{3}{4}\right)^{-1/2-k/2} \log \left(n+\frac{l}{2} + \frac{3}{4}\right)
    \right\} 
\end{equation}
into Eq.~(\ref{eq_1f}) and Taylor expanding the WKB condition in 
$1/(n+l/2+3/4) \ll 1$ one obtains to leading order
\begin{eqnarray}
\fl
  E_{n,l} &\approx& \frac{6 {\overline{m}}^2 + 3 {\mu}^2 \pi + 2l{\mu}^2\pi + 
    4{\mu}^2 n\pi - 2{(\Delta m)}^2 \log (\overline{m}\mu) - 
    {(\Delta m)}^2\left( -2 + \log (16\pi ) \right) }{\mu
    {\sqrt{3 + 2l + 4n}}{\sqrt{\pi}}}\nonumber \\
\fl
  && + \frac{{(\Delta m)}^2\left( 4\overline{m}{\sqrt{3 + 2l + 4n}} - 
    \mu\left( 3 + 2l + 4n \right){\sqrt{\pi }} \right)
    \log (\frac{3}{4} + \frac{l}{2} + n)}{{\mu}^2
    {\left( 3 + 2l + 4n \right) }^{\frac{3}{2}}\pi } + \dots .
\label{eq_1g}
\end{eqnarray}

The dashed curve of Fig.~\ref{fig:error_Em1eqm2} displays the error over the
spectrum obtained using eq.~(\ref{eq_1g}), showing that accurate predictions 
can be obtained with this simple formula.

\section{Conclusions}
\label{sec:concl}

In this paper we have applied to the integrals occurring in the application 
of the WKB a simple method, which allows one to express them analytically 
and to arbitrary precision in terms of elementary functions (a formal proof of
the convergence of such method has been previously given in Ref.~\cite{AS:04}).
This procedure is advantageous even in the case (not so frequent) in which 
the integral can be performed analytically, since the exact result will in
general contain special functions, which are usualy more difficult to handle. 

Successively, the approximate WKB quantization condition obtained in the 
first stage has been inverted, thus providing analytical formulas for the
spectrum of a particle in a given potential. 

In this paper we have considered two physical applications, which are relevant
in meson spectroscopy, showing that in both cases excellent accuracy is
obtained. The precision of our results essentially depends upon the intrinsic 
precision of the leading order WKB approximation, since the WKB integrals
can be approximated analytically to arbitrary precision using our method.
As a matter of fact we have also applied our method to one-dimensional
anharmonic potentials (not discussed here) --- using a higher order WKB
calculation --- and obtaining highly accurate formulas for the spectrum.
Given the simplicity of our method we believe that it can be applied to a wide
class of problems.

\section{Acknowledgments}

P.A. acknowledges support of Conacyt grant no. C01-40633/A-1.

\appendix

\section{}

In this Appendix we provide the explicit expressions of the turning points and
of the minimum of the Cornell potential (\ref{eq:Corneff}) and also the
functions ${\mathcal H}_{0\pm}$, ${\mathcal H}_{1\pm}^a$,
${\mathcal H}_{1\pm}^b$ and ${\mathcal H}_{1\pm}^c$ introduced in
Sect.~\ref{subsec:Cornell}. 

For the turning points we follow Ref.~\cite{Brau00}:
\numparts
\begin{eqnarray}
\fl
  r_{+} = \frac{1}{3a} \left\{ E + 2\sqrt{E^2 + 3a\kappa} \cos \left[ 
    \frac{1}{3} {\rm sec}^{-1} \left[ 
    \frac{4\mu (E^2+3a\kappa)^{3/2}}{4E^3\mu + 9a(-3aL^2+2E\kappa\mu)} \right]
    \right] \right\} , \\
\fl
  r_{-} = \frac{1}{3a} \left\{ E + 2\sqrt{E^2 + 3a\kappa} \cos \left[ 
    \frac{4\pi}{3} + \frac{1}{3} {\rm sec}^{-1} \left[ 
    \frac{4\mu (E^2+3a\kappa)^{3/2}}{4E^3\mu + 9a(-3aL^2+2E\kappa\mu)} \right]
    \right] \right\}. \nonumber \\
\end{eqnarray}
\endnumparts
The position of the minimun of the effective Cornel potential is:
\begin{equation}
  r_0 = \frac{\sqrt[3]{2} \left(9 a^2 L^2 \mu ^2+\sqrt{3} \sqrt{a^3 \mu ^4
    \left(27 a L^4+4 k^3 \mu ^2\right)}\right)^{2/3}-2 \sqrt[3]{3} a k \mu
    ^2}{6^{2/3} a \mu \sqrt[3]{9 a^2 L^2 \mu ^2+\sqrt{3} \sqrt{a^3 \mu ^4
    \left(27 a L^4+4 k^3 \mu ^2\right)}}} .
\end{equation}
The functions ${\mathcal H}$ entering the first order approximation to the WKB
integral of Eq.~(\ref{eq:J1pm}) are defined as follows:
\numparts
\begin{eqnarray}
\fl
  {\mathcal H}_{0\pm} &=& \pm \int_{r_0}^{r_{\pm}} dr 
    \sqrt{(r_{\pm}-r)(r+r_{\pm}-2r_0)} = \frac{\pi}{4}(r_{\pm}-r_0)^2 \\
\fl
  {\mathcal H}_{1\pm}^a &=& \pm \int_{r_0}^{r_{\pm}} dr 
    \sqrt{\frac{r_{\pm}-r}{r+r_{\pm}-2r_0}} = 
    \pm \left(\frac{\pi}{2}-1\right) (r_{\pm}-r_0) \\ 
\fl
  {\mathcal H}_{1+}^b &=& \int_{r_0}^{r_{+}} dr 
    \sqrt{\frac{r_{+}-r}{r+r_{+}-2r_0}}\frac{1}{r} = 
    -\frac{\pi}{2} + 
    \frac{1}{\sqrt{1-\frac{\displaystyle 2 r_0}{\displaystyle r_{+}}}}
    \log\left[\frac{r_{+}}{r_0} \left( 1 + 
    \sqrt{1-\frac{\displaystyle 2 r_0}{\displaystyle r_{+}}}\right)-1\right] 
    \nonumber \\
\fl
  {\mathcal H}_{1-}^b &=& \int_{r_{-}}^{r_0} dr 
    \sqrt{\frac{r_{-}-r}{r+r_{-}-2r_0}}\frac{1}{r} = 
    \frac{\pi}{2} - 
    \frac{2}{\sqrt{\frac{\displaystyle 2 r_0}{\displaystyle r_{-}}-1}}
    \arctan
    \sqrt{\frac{\displaystyle 2 r_0}{\displaystyle r_{-}}-1} \\
\fl
  {\mathcal H}_{1+}^c &=& \int_{r_0}^{r_{+}} dr 
    \sqrt{\frac{r_{+}-r}{r+r_{+}-2r_0}}\frac{1}{r^2} \nonumber \\
\fl
  &=& \frac{r_{+}-r_0}{ r_{+}-2r_0} \left\{\frac{1}{r_0} - \frac{1}{r_{+}}
    \frac{1}{\sqrt{1-\frac{\displaystyle 2 r_0}{\displaystyle r_{+}}}}
    \log\left[\frac{r_{+}}{r_0} \left( 1 + 
    \sqrt{1-\frac{\displaystyle 2 r_0}{\displaystyle r_{+}}}\right)-1\right]
    \right\} \\
\fl
  {\mathcal H}_{1-}^c &=& \int_{r_{-}}^{r_0} dr 
    \sqrt{\frac{r_{-}-r}{r+r_{-}-2r_0}}\frac{1}{r^2} = 
    \frac{r_{-}-r_0}{ r_{-}-2r_0} \left[-\frac{1}{r_0} + \frac{2}{r_{-}}
    \frac{1}{\sqrt{\frac{\displaystyle 2 r_0}{\displaystyle r_{-}}-1}}
    \arctan
    \sqrt{\frac{\displaystyle 2 r_0}{\displaystyle r_{-}}-1} \right]
    \nonumber \\
\fl
\end{eqnarray}
\endnumparts

\end{document}